\documentclass[journal]{IEEEtran}
\usepackage[numbers,sort&compress,square]{natbib}
\usepackage[font=footnotesize]{caption}
\usepackage{setspace}
\usepackage{amsmath}
\usepackage{nccmath}
\usepackage{amsfonts}
\usepackage{amsthm}
\usepackage{amssymb}
\usepackage{relsize}

\usepackage{multirow}
\usepackage{stackengine}

\usepackage{float}
\usepackage{ifthen}
\usepackage{etoolbox}
\usepackage{soul}
\usepackage[binary-units=true]{siunitx}

\usepackage{color}
\usepackage[off]{auto-pst-pdf}
\usepackage{array}
\usepackage{psfrag}
\usepackage{epstopdf}
\usepackage{epsfig}
\usepackage[prefix]{nomencl}

\usepackage{graphicx}

\usepackage[]{algorithmicx}
\usepackage{algpseudocode}
\usepackage{algorithm}
\usepackage{booktabs}
\usepackage{enumerate}
\usepackage{enumitem}

\usepackage[caption=false, font=footnotesize]{subfig}
\algnewcommand{\LeftComment}[1]{\Statex \(\triangleright\) #1}

\usepackage[mathscr]{euscript}
 \let\mathscr\relax% just so we can load this and rsfs
\usepackage[scr]{rsfso}

\usepackage{tikz}
\usetikzlibrary{calc,shapes,decorations,decorations.text, mindmap,shadings,patterns,matrix,arrows,intersections,automata,backgrounds, arrows.meta}
\usepackage{tikz}
\usetikzlibrary{calc,shapes,decorations,decorations.text, mindmap,shadings,patterns,matrix,arrows,intersections,automata,backgrounds, arrows.meta}
\usetikzlibrary{calc}
\usepackage[american,siunitx]{circuitikz}

\usetikzlibrary{positioning}
\usepackage{scalefnt}

\usetikzlibrary{arrows.meta,
	decorations.pathreplacing,
	positioning,
	shadows
}
\usepackage{enumitem}
\setlist{nosep,leftmargin=*}

\usepackage[american,siunitx]{circuitikz}

\theoremstyle{definition}
\newtheorem{definition}{Definition}[section]
\theoremstyle{remark}

\usepackage{tocbibind} % This adds Bibiliography in to the table of content.
\usepackage[version=3]{mhchem} 
\usepackage[utf8]{inputenc}
\usepackage{datetime}
\usepackage{textcomp}
\usepackage{longtable}
\usepackage{amssymb}
\usepackage{cleveref}
\usepackage{tabularx}
\usepackage{type1cm}
\usepackage{lettrine}

\allowdisplaybreaks

\ifCLASSINFOpdf
\else
\fi
\begin{document}
	
%\setlength{\abovedisplayskip}{3pt}
%\setlength{\belowdisplayskip}{3pt}
\title{A Turvey-Shapley Value Method for Distribution Network Cost Allocation}


% \author{\IEEEauthorblockN{Donald Azuatalam, \textit{Student MIEEE}, Archie Chapman, \textit{MIEEE} and Gregor Verbi\v{c}, \textit{Senior MIEEE}}
% \IEEEauthorblockA{School of Electrical and Information Engineering, The University of Sydney, Sydney, Australia\\
% 	%	\IEEEauthorblockA{
% 		Email:\{donald.azuatalam, archie.chapman, gregor.verbic\}@sydney.edu.au}}

\author{Donald~Azuatalam,~\IEEEmembership{Graduate Student Member,~IEEE,}
	Archie~C.~Chapman,~\IEEEmembership{Member,~IEEE,}
	and~Gregor~Verbi\v{c},~\IEEEmembership{Senior Member,~IEEE}% <-this % stops a space
	\thanks{Donald Azuatalam, Archie C. Chapman and Gregor Verbi\v{c}, are with the School of Electrical and Information Engineering, The University of Sydney, Sydney, New South Wales, Australia. E-mail: donald.azuatalam@sydney.edu.au, archie.chapman@sydney.edu.au, gregor.verbic@sydney.edu.au.}}

\maketitle

\begin{abstract}
This paper proposes a novel cost-reflective and computationally efficient method for allocating distribution network costs to residential customers. First, the method estimates the growth in peak demand with a 50\% probability of exceedance (50POE) and the associated network augmentation costs using a probabilistic long-run marginal cost computation based on the Turvey perturbation method. 
Second, it allocates these costs to customers on a cost-causal basis using the Shapley value solution concept. 
To overcome the intractability of the exact Shapley value computation for real-world applications, we implement a fast, scalable and efficient clustering technique based on customers’ peak demand contribution, which drastically reduces the Shapley value computation time. 
Using customer load traces from an Australian smart grid trial (\textit{Solar Home Electricity Data}), we demonstrate the efficacy of our method by comparing it with established energy- and peak demand-based cost allocation approaches.
\end{abstract}

% This paper proposes a novel, cost-reflective and computationally efficient methodology for allocating distribution network costs to residential customers based on a probabilistic \textit{Turvey} long-run marginal cost (LRMC) method and the \textit{Shapley value} game-theoretic solution concept. 
% Since capacity network costs are forward-looking, we use the {Turvey} method to estimate the 50\% %POE
% probability-of-exceedence growth in peak demand and the associated augmentation cost. 
% We then allocate these costs to customers on a cost-causal basis by computing their {Shapley value}. 
% In light of the intractability of the exact Shapley value computation for large number of players, which is the case in many real-world applications, recent work in this area have used various sampling approaches to reduce the computational burden. However, even with these methods, the estimation error and/or the computation time increases with a significant number of players. 
% To overcome these drawbacks, we implement a fast, scalable and efficient clustering technique based on customers' peak demand contribution that %, in conjunction with the Turvey-Shapley value method to allocate network costs, 
% drastically reduces the Shapley value computation time. 
% Using customer load traces from the Australian \textit{Solar Home Electricity Data}, our numerical results demonstrate that the Shapley value allocates cost in the most cost-reflective manner regardless of PV adoption, compared to other cost allocation methodologies. 

%efficacy of our methodology, showing that
\begin{IEEEkeywords}
Turvey perturbation method, 
long-run marginal cost, 
Shapley value, 
k-means clustering, 
cost-causality, 
demand-based tariffs, 
cost-reflective network pricing
\end{IEEEkeywords}

% \fi
\IEEEpeerreviewmaketitle

\section{Introduction} \label{section1}
The rapid rise in the penetration levels of distributed energy resources in low-voltage distribution networks necessitates the design of network tariffs that allocate associated network costs in an efficient, fair and equitable manner to network users. Hence, distribution network service providers (DNSPs) and regulators in most jurisdictions are challenged with the tasks of designing efficient tariffs that are reflective of network cost drivers \cite{picciariello2015distributed}. Recent studies in this area have explored different methods for distribution network pricing,  including transmission network pricing methodologies, such as  \textit{Locational
Marginal Pricing} (LMP),
\textit{Postage Stamp} (PS), \textit{MW-Mile}, \textit{MVA-Mile}, \textit{Average
Participation} (AP), and \textit{Marginal Participation} (MP) \cite{rubio2000marginal,zolezzi2001review,sotkiewicz2006allocation,li2008cost,brown2015efficient}, \textit{Ramsey-Boiteux pricing}, \textit{cooperative game theory} or other extemporaneous cost allocation methodologies. 
However, in order to establish the performance of these methods with respect to established tariff design principles \cite{bonbright1961principles}, which includes \textit{cost reflectiveness}, \textit{efficiency}, \textit{stability} and \textit{fairness}, we need to define a measure (benchmark) of overall performance, with which to compare existing methods. To this end, the purpose of our study is to use a principled cost allocation method as a performance benchmark for other allocation methodologies. %the actual structure of the tariff needs to be explored too.
% \textit{Ramsey Pricing},

%One important tariff design goal is to recover both the residual and long-run marginal cost components of providing network services, and for this goal to be achieved,

%whether a tariff be two- or three-part, 
%efficiently and appropriately recover these cost components. 

\par Distribution network costs typically comprises three major cost components--energy costs, long-run marginal cost (LRMC), and residual (such as retail charges) costs. In order to adequately recover these cost components, network tariffs should be structured such that the fixed, energy and/or demand charges efficiently send the right price signals to customers to respond appropriately.
For example, \cite{abdelmotteleb2018designing} used a three-part tariff for distribution network cost allocation. Here, the residual costs (\$/day) were recovered through \textit{Ramsey pricing}, while the LRMC (\$/kW) and energy costs (\$/kWh) were recovered through \textit{coincident peak pricing} and \textit{distribution locational marginal pricing} (DLMP) respectively. 

\par Nevertheless, network tariffs historically only consisted of energy-based (volumetric) and residual charges due to two reasons: (i) there was little need to signal the LRMCs because loads per feeder were relatively flat and (ii) pricing mechanisms available to utilities were severely constrained by metering technology. However, with the introduction of smart meters, it is possible to implement tariffs which reflect congestion costs that drives network investments. As such, network tariffs should be based on customers demand at network peak \cite{lewis1941two,boiteux1952determination,nijhuis2017analysis,passey2017designing,abdelmotteleb2018designing,nelson2013new}. It should be time- and location-specific and should account for network losses and actual customer energy use. Additionally, a fair and equitable tariff should also eliminate or reduce inter-customer subsidies created due to PV owners, while safeguarding vulnerable customers \cite{picciariello2015electricity,simshauser2014network}.

%the need to signal LRMCs were less
%access to energy was seen as a developmental goal and a public good, so government subsidy of electricity was generally accepted.
%a lack in understanding of tariff design and cost drivers  in  the  electricity  industry, and the unavailability of smart meters and

\par Unlike volumetric tariffs, peak demand-based tariffs are robust to technological changes (such as solar PV, EV or battery storage) which reshapes customers' demand profiles while effectively signalling peak demand costs to customers \cite{simshauser2014network,pimm2018time}. Thus, residual and/or LRMC costs can be recovered partly through demand-charges instead of constantly increasing fixed or energy charges for all customers \cite{brown2015efficient,ahmad2018pricing}. 
So far, \textit{coincident peak pricing} (CPP) and \textit{critical peak pricing} have been proposed to mitigate the impacts of DER on the equity of network cost allocations. Although peak demand-based tariffs are more complex than energy-based tariffs, they can better allocate network costs on a cost-causal basis and ensure a stable revenue for network companies \cite{simshauser2014network,ahmad2018pricing}. 
Furthermore, \cite{ahmad2018rate} showed that customer bill volatility reduces with demand-based tariffs compared to \textit{real time pricing} and \textit{time-of-use tariffs}. Contrarily, \cite{passey2017designing} tested {demand-based tariffs} proposed by the Australian Energy Regulator (AER) on households in Sydney. It was concluded that without due adjustments made, these tariffs are low in cost-reflectivity. Generally, the suitability of network tariffs in terms of fairness and cost-reflectivity depends on the assumptions made in the tariff design and on customers' price response \cite{stenner2015australian}. This further highlights the need for a principled cost allocation benchmark.

\par Further still, the distribution network tariff design problem encompasses more than the aforementioned requirements. Beyond these, there are other three salient tariff design questions that need to be answered in order to achieve a cost-reflective pricing and appropriate customer response: 
(i) What LRMC calculation method should be used -- the \textit{Turvey perturbation method} or the \textit{Average Incremental Cost method}? 
(ii) What peak demand should be the basis for charging customers -- individual customer peak, distribution network coincident peak or zone-substation peak? and,
(iii) What is the optimal frequency of peak demand measurement -- monthly or yearly basis? \cite{ahmad2018pricing}. The answers to these questions form the basis for practically implementing tariffs that better recuperate forward-looking network costs. However, there are no clear-cut answers to these questions, because choices for network companies depend on other factors, such as customer socio-demographics, customer class, and the availability of smart meters and energy management systems.  
Nonetheless, recent research in the area argues that demand-based tariffs should be based on network coincident peak since it better signals LRMC. 
However, in practice, customers' coincident peak demand is hard to measure and thus CPP is difficult to implement. Thus, demand charges are a step-forward to attaining optimal network tariffs.

\par In this paper, we seek, based on established economic principles, to make a further step towards the design of equitable network pricing. Our focus is to provide a measure for the fair and efficient allocation of costs that signal the drivers of future network investment. 
To achieve this, we develop a novel method to  apportion the LRMC, using a probabilistic approach  to the \textit{Turvey perturbation method} (\cite{turvey1969marginal}) linked via the \textit{characteristic function} of a cooperative game\footnote{A cooperative game models a game where a group of players cooperate to earn a joint reward, which has to be shared among the players in a fair and stable way.} to the \textit{Shapley value} (SV) cost allocation rule \cite{Shapley1953,ChalkiadakisEtal2011}. 

In more detail, the {Turvey perturbation method} is a forward-looking and more time- and location-specific method for LRMC estimation, compared to the simpler {average incremental cost} methods widely used by network companies \cite{ahmad2018pricing}. 
Furthermore, \cite{biggardarr,neraconsult} argue that the {Turvey perturbation} method is the preferred option since it better aligns with the underpinning principles governing LRMC. However, research in \cite{toothrichard} concluded that both methods can be equally used for LRMC calculations.

%\par Turvey method toothrichard,neraconsult,biggardarr} 

%\cite{biggardarr} %Turvey versus AIC
% Actually, the December 2011 proposes two approaches to computing LRMC: The perturbation or Turvey
% approach and the Average Incremental Cost or AIC approach. But the paper expresses a clear preference for the
% former: “In our opinion the perturbation approach should be preferred over an AIC approach because it most
% closely aligns with the principles underpinning the concept of LRMC”.

%\cite{ahmadpricing}
% What should be the basis of the estimation of LRMC. The principles allow for the use of either the Average Incremental Cost or Turvey (Perturbation) methods. The AIC approach is simpler, is more widely used in the DNSPs, but is less time or location specific than the Turvey method. In contrast, the Turvey method can provide a stronger locational signal and is more sensitive to the timing of new investment requirements. Hence, while the AIC may be preferred in estimating variable rates for the standard tariffs, the Turvey method may be more appropriate for locational price signals.

%ghassemi2008cooperative, 
\par At the same time, the SV gives a vector-valued solution to a cooperative \textit{transferable utility} (TU) game, where the total cost or worth of a coalition is defined by a single-valued \textit{characteristic function}.
In our method, the characteristic function is the probabilistic LRMC defined using the Turvey perturbation method. 
The SV has found several applications including consumer demand response compensation \cite{o2015shapley,bakr2015using,Chapman_IREP2017}, transmission network cost allocation \cite{zolezzi2002transmission,tan2002application,khare2015shapley}, distribution network loss allocation \cite{sharma2017loss}, and other cost allocation problems \cite{ghassemi2008cooperative,stanojevic2010economic,byun2009fair,o2015shapley,bakr2015using}.
However, due to the computational complexity of computing the \textit{exact} SV for a large number of players, it's application is usually limited to small problems. Recent research in this area, nonetheless, has seen developments of approximate methods of calculating the SV in polynomial time \cite{fatima2008linear,fatima2007randomized,stanojevic2010economic,david2005shapley,castro2009polynomial,byun2009fair,bakr2015using}. %owen1972multilinear,
In \cite{bakr2015using}, a comparison of the accuracy and scalability of two approximate SV computation algorithms was made, namely \textit{linear-time approximation} \cite{fatima2008linear} and \textit{stratified sampling} \cite{castro2009polynomial,maleki2013bounding} techniques. While the \textit{stratified sampling} approach was more accurate, the \textit{linear-time approximation} required less memory and computation time as the number of players increased. This is a general finding, so with these methods, there is always a trade-off between accuracy and computational complexity. Moreover, some of these methods are only suited to \textit{weighted voting games}, which does not match our cost-allocation problem. 
In a different research direction, a clustering approach was adopted in \cite{david2005shapley}, where customers are segmented into major classes. However, customers in the same class are assigned the same SV. Conversely, for distribution network cost allocation, this is not the case, as the SV should be different for all customers.

%including exact -- \textit{generating function} \cite{shapley1962values}
% \par Shapley value computation algorithms randomised \cite{fatima2008linear,stanojevic2010economic}
% Approximating Optimum Stratified Sampling \cite{david2005shapley}
% sampling \cite{castro2009polynomial}
% clustering in groups \cite{david2005shapley}
% sampling error reduction \cite{maleki2013bounding}
\par In light of these shortcomings, we derive a computationally-efficient clustering algorithm, to allocate network costs based on the Turvey-Shapley value method. The SV is computed at the level of clusters, and individual customers are allotted a portion of this SV based on their average coincident peak demand contribution to each coalition of their representative cluster. This approximation approach is validated by comparison to the exact SV calculation, for which the SV estimation error is shown to be small and reduces as the number of customers increases (i.e.~when approximation become computationally necessary). 
Furthermore, as the SV method best allocates network costs in a principled, fair and stable manner, we used it as a benchmark to measure the cost-reflectivity of other cost allocation methodologies. 
In summary, the analysis in this paper extends the preliminary results in our earlier conference paper \cite{azuatalam2019shapley} in the following ways: 

%compared the SV cost allocation with other peak demand cost allocation methodologies, which showed that the 

\begin{itemize}
	\item We propose a probabilistic approach to the {Turvey} LRMC  computation via a Weibull distribution, which gives an unbiased estimate of forward-looking network costs. 
	\item We propose a peak load contribution clustering technique interleaved with the Turvey-Shapley value method to compute the Shapley value for large number of customers with low computation time and estimation error.
	\item We demonstrate the effectiveness of our methodology using real customer load traces from the \textit{Solar Home Electricity Data}\footnote{Dataset is available at https://data.nsw.gov.au/}. Our results show that the proposed allocation method is the most reflective of network capacity costs compared to other cost allocation methodologies.
\end{itemize}

\par The remainder of this paper is structured as follows. Section \ref{prelim} provides preliminaries on cooperative game theory and the Shapley value. 
In Section \ref{method}, we describe the methodology, including the Turvey pricing method. The results are presented and discussed in Section \ref{results}. 
Section \ref{conclusions} concludes.

%theory
% \cite{Shapley1953,peleg2007introduction,winter2002shapley,david2005shapley,winter1989value,hart1988potential}
% %application and approximation of shapley value
% \cite{ghassemi2008cooperative,stanojevic2010economic,fatima2008linear,byun2009fair,zolezzi2002transmission,fatima2007randomized,tan2002application}

% \par Aside \cite{bonbright1961principles}, other authors \cite{nelson1964marginal,turvey1964marginal,houthakker1951electricity} have also identified that good tariff design should be based on marginal pricing. 
%  The major network investment cost driver is aggregate network peak demand
% By so doing, tariffs will be designed so that customers are charged according to the extent to which they contribute to network investment cost drivers.

\vspace{-0.1em}

\section{Preliminaries} \label{prelim}
In this section, we provide a background to cooperative games, the \textit{Shapley value} and its characteristics, and the \textit{Turvey perturbation method}. %\cite{Schmeidler1969,vonNeumannMorgenstern1944}
%\subsection{Preliminaries}

% We begin with some basic terminology,
% which can be found in most game theory texts covering cooperative games (e.g.~\cite{OsbourneRubinstein1994}); 
% the foundations of cooperative games are laid out in \cite{vonNeumannMorgenstern1944}.  

% If the players in a cooperative game agree to work together, 
% they form a \emph{coalition}. 
% If all $n$ players form a coalition, it is called the \emph{grand coalition}.
% Each player incurs some private \emph{cost} in completing its component of the joint action, while collectively, 
% the joint action has some \emph{worth} associated with it (i.e. revenue). 
% The difference between the sum of the players costs' and the worth is called the coalition's \emph{surplus}.
% The task is to divide the surplus in such a way to ensure that the coalition is \emph{stable}. 
% This means that no other smaller coalition could \emph{deviate} to form a new coalition that can also complete the joint action, 
% but divides its surplus in a way that (weakly) improves the allocation to all players in the new coalition.
% Beyond this broad and loose definition, 
% many refinements to the permissible set of deviations and desirable characteristics of the final surplus division have been proposed, 
% giving rise to a surfeit of \emph{solution concepts}. 

%~\cite{OsbourneRubinstein1994,vonNeumannMorgenstern1944}

\subsection{Cooperative Games} 
Formally, we consider the class of \emph{transferable utility} (TU) games, 
which are cooperative games that allow the transfer of worths between players. 
If the players in a cooperative game agree to work together, 
they form a \emph{coalition}. If all $n$ players form a coalition, it is called the \emph{grand coalition}.
Each player incurs some private \emph{cost} in completing its component of the joint action, while collectively, 
the joint action has some \emph{worth} associated with it. 

%(cf. non-transferable utility games).  

\begin{definition}
	A TU game is given by $\Gamma = \langle \mathcal{N},w\rangle$ where: 
	\begin{itemize}
		\item $\mathcal{N}$ is a set of $n$ \emph{players}, and 
		\item  $w(\mathcal{S})$ is a \emph{characteristic function}, $w:2^{n}\to \Re_+$ with $w(\emptyset) = 0$, 
		that maps from each possible coalition $\mathcal{S}\subseteq \mathcal{N}$ to the \emph{worth} of $\mathcal{S}$.
	\end{itemize}
\end{definition}
% It is important to note that the characteristic function implicitly assumes that the players follow an \emph{optimal} joint action --- 
% and in practice computing this optimal action can be difficult.

%Solution concepts in cooperative game theory define divisions of the group reward among players, while considering the rewards available to each alternative coalition of players.
%\subsection{Cooperative Solution Concepts}\label{sec:cooperativeSolutions}
 Before defining the Shapley value, we first formally define some important characteristics of any solution to a TU game.

\begin{definition}
	Given a $\Gamma$, a solution concept defines a \emph{worth} to each player, which is a vector of transfers (worths), 
	$\phi = (\phi_1,\ldots, \phi_i,\ldots,\phi_n) \in \Re^n$. %Where $\Re$ is a set of real numbers.
\end{definition}

We denote the sum of worths as $\sum_{i\in \mathcal{S}} \phi_i = \phi(\mathcal{S})$. 
Some desirable properties of solutions concepts include the following; a solution is:
\begin{itemize}
% 	\item \emph{Feasible} if $\phi(\mathcal{S}) \leq w(\mathcal{S})$, meaning the total of all the worths is less than the coalition's worth,
	\item \emph{Efficient} if $\phi(\mathcal{S}) = w(\mathcal{S})$, so that the worth vector exactly divides the coalitions worth,
% 	\item \emph{Individually rational} if $t_i \geq w(i)$ for all $i \in N$, meaning the worth of a player is at least what it can get by acting alone. 
	\item \emph{Symmetric} if $\phi_i = \phi_j$ if $w(\mathcal{S}\cup \{i\}) = w(\mathcal{S}\cup \{j\}),\ \forall ~\mathcal{S}\subseteq N\setminus \{i,j\}$. This means that equal worths are made to symmetric players, where symmetry means that we can exchange one player for the other in any coalition that contains only one of the players and not change the coalition's worth. %This property is also called \emph{anonymity}, because the players' labels do not affect their worth.
	\item \emph{Additive} if for any two additive games % with characteristic functions $w_1$ and $w_2$, the game $\langle N, w_1 + w_2 \rangle$ is additive if $[w_1 + w_2](S) = w_1(S) + w_2(S)$ for any coalition $S\subseteq N$. 
	the solution can be given by
	$\phi_i(v_1+v_2) = \phi_i(v_1) + \phi_i(v_2)$ for all players.
	That is, an additive solution assigns worths to the players in the combined game that are the sum of their worths in the two individual games.
	\item \emph{Zero worth to a null player} if a player $i$ that contributes nothing to any coalition, such that $w(\mathcal{S}\cup \{i\})=w(\mathcal{S})$ for all $\mathcal{S}$, then the player receives a worth of 0.
\end{itemize}

\subsection{The Shapley Value (SV)} \label{section_shapley_value}
Solution concepts in cooperative game theory define divisions of the group reward among players, 
while considering the rewards available to each alternative coalition of players. The SV is one of such solution concepts which also satisfies the desirable properties listed above, by virtue of its definition.

\begin{definition}
	The SV allocates to player $i$ in a coalitional game $\langle w,\mathcal{N}\rangle$ the worth:
\begin{equation} \label{defShapleyValue1}
\phi _{i}(w)
= \frac {1}{n}\sum_{\mathcal{S}\subseteq \mathcal{N}\setminus i} 
\binom{n-1}{|\mathcal{S}|}^{-1}\left(w(\mathcal{S}\cup \{i\})-w(\mathcal{S})\right)
\end{equation}	
\end{definition}

% 	The SV can also be written as in \eqref{defShapleyValue2}
% 	\begin{equation}\label{defShapleyValue2}
% 	\phi_{i}(w) = \sum_{\substack{\mathcal{S}\subseteq \mathcal{N} \\ i \in \mathcal{S}}}
% 	\frac{(|\mathcal{S}|-1)!\;(n-|\mathcal{S}|)!}{n!}\, \left(w(\mathcal{S}) - w(\mathcal{S} - \{i\})\right)
% 	\end{equation}

% 	\begin{equation}\label{defShapleyValue1}
% 	\phi_{i}(w) =  \sum_{\mathcal{S}\subseteq \mathcal{N} \setminus \{i\}}
% 	\frac{|S|!\;(n-|S|-1)!}{n!}\, \left(w(\mathcal{S}\cup \{i\}) - w(\mathcal{S})\right)
% 	\end{equation}

Here, the vector-valued function $\phi$ has the following intuitive interpretation:
consider a coalition being formed by adding one player at a time. 
When $i$ joins the coalition $\mathcal{S}$, its \emph{marginal worth} is given by $w(\mathcal{S}\cup \{i\}) - w(\mathcal{S})$. Then, for each player, its SV worth is the average of its marginal contributions over the possible different orders in which the coalition can be formed. %Alternatively, the SV can be rearranged as:
%This is the last part of the expression in \eqref{defShapleyValue1} above. 
%and expressed 

%\section{Methodology} 
%\subsection{Approximating the Shapley value} \label{section3}

The expression in \eqref{defShapleyValue1} can also be interpreted as a player's contribution to all subsets of $\mathcal{N}$ that do not contain it, where the binomial term is the number of coalitions of size $|\mathcal{S}|$. 
We can further expand this expression to identify a useful approximation.
Specifically, the summation in \eqref{defShapleyValue1} can be expressed in terms of the size of the coalitions 
% sans $i$, $S\subseteq N\setminus \{i\}$, 
that $i$ is added to, as follows:
\begin{equation}\label{eqn:SV_expanded}
\phi_{i}(w)
= \frac {1}{n}\sum_{k=0}^{n-1}
\left( \binom{n-1}{k}^{-1} 
\sum_{\mathcal{S}\in \mathcal{S}^k}  \left(w(\mathcal{S}\cup \{i\}) - w(\mathcal{S})\right) \right)
\end{equation}
where: $\mathcal{S}^k = \{\mathcal{S}\subseteq \mathcal{N}\setminus i \,:\, |\mathcal{S}| = k \}$
is the set of coalitions of size $k$ that exclude $i$.
We can now approach the SV by approximating the inner term for each size $k \in \{0,\ldots,n-1\}$.

One approach is to statistically estimate the term:
\begin{equation} \label{eqn:SV_randomised}
\chi^k = \binom{n-1}{k}^{-1} \sum_{\mathcal{S}\in \mathcal{S}^k}  \left(w(\mathcal{S}\cup \{i\}) - w(\mathcal{S})\right)
\end{equation}
using a sample-based approach. This is a randomised sampling algorithm described in detail in Section \ref{section_randomised_sampling}.

\begin{algorithm}[t]
	\caption{Exact Turvey-Shapley Value Algorithm (\textit{Exact})} \label{ShapleyvalueExact1}
	\small
	%\Procedure{MyProcedure}{}
	
	%\textbf{Inputs:}\\
		$\mathcal{N}$: set of players/customers, $\mathcal{N} = \{1,..,n\}$ \\
	$\mathcal{L}$: set of all coalitions, $\mathcal{L} = \{\mathcal{S}^1,..,\mathcal{S}^l,..., \mathcal{S}^{2^n - 1}$\}
	% \left\vert\mathcal{L}\right\vert=
\begin{algorithmic}[1]
%\State \textbf{Compute} coalition matrix
%\State \textbf{Compute} cost for each coalition $\mathcal{L}^l \in \mathcal{L}$
		\For{each customer $i \in \mathcal{N}$}
		\For{each coalition $\mathcal{S} \in \mathcal{L}:i \in \mathcal{S}$}
		\State $\mu$ = 50POE yearly peak demand of $\mathcal{S}$
		\State \textbf{Find} $X\sim\mathcal{W}(\alpha,\beta)$ of $\mu$ exceeding line limit $x$
		
		\State $\alpha = \mu/ \Gamma(1+1/\beta)$ with $\beta=1.5$ 
		\If{$P(X \geq x) < 0.001$}
				\State $w(\mathcal{S}) = P(X \geq x)Y$
		\Else
                \State $w(\mathcal{S}) = 0$
		\EndIf
		\State \textbf{Compute} $M_{\mathcal{S}_i}=w(\mathcal{S}) - w(\mathcal{S} - \{i\})$ 
		\State \textbf{Compute} $K_{\mathcal{S}_i}=(1/n!)(|\mathcal{S}|-1)!\;(n-|\mathcal{S}|)!$
		\EndFor
		\State \textbf{Return} SV of $i$, $\phi_{i}(w) = \sum_{\substack{\mathcal{S}\in \mathcal{L}}} K_{\mathcal{S}_i} M_{\mathcal{S}_i}$
		\EndFor
	\end{algorithmic}
\end{algorithm}

\subsection{The Turvey Perturbation Method} \label{turvey_method_section}
The \textit{Turvey perturbation method} \cite{turvey1969marginal} is one technique used to estimate the LRMC of capacity-based investments. It quantifies the effects of a (small) permanent change in demand $Q$ on future capital costs $C$. It is defined in \cite{toothrichard} as:

% \begin{equation} \label{eqn:turvey}
% \textrm{Turvey LRMC} = \frac{\partial C}{\partial t} \bigg/ \frac{\partial Q}{\partial t}  = \frac{\partial C}{\partial Q} 
% \end{equation}

\begin{equation} \label{eqn:turvey}
\textrm{Turvey LRMC}   = \frac{PV(\Delta C)}{PV(\Delta Q)} 
\end{equation}

%= \frac{\partial C}{\partial t}  \frac{\partial t}{\partial Q}

The expression in \eqref{eqn:turvey} translates to--the ratio of the present value of change in costs (due to a permanent change in demand) to the present value of the permanent change in demand. The Turvey perturbation method therefore involves forecasting demand over the estimation period, with a certain confidence level. For this, we assume a small growth in yearly peak demand with a 50\% probability of exceedance. 

As explained in detail in the next section, in our methodology, the probability that this value exceeds the network line limit informs the LRMC. 
This probabilistic approach to the Turvey perturbation method, achieved via a Weibull distribution, is used to construct the \textit{characteristic function} for the SV computation.

\section{Methodology} \label{method}

In this section, we detail the steps taken to assess the cost-reflectivity in the allocation of distribution network tariffs to LV residential customers, with the SV allocation being the benchmark. First, we explain the \textit{Turvey-Shapley value} LRMC estimation and allocation methodology. Second, we describe algorithms to determine the exact SV for a set of customers $\mathcal{N}$, and its approximation. The approximation algorithms are required to compute the SV for $n>25$ players, with lower computational burden and minimal loss in accuracy. 

%s explained in \ref{turvey_method_section}, interleaving a probabilistic approach to the \textit{Turvey method} with the SV \textit{characteristic function} is a novel methodological contribution. Algorithm \ref{ShapleyvalueExact1} illustrates the Turvey-Shapley value methodology, which computes the exact SV for $n$ customers. 

%\vspace{-0.5em}

\subsection{Turvey-Shapley Value LRMC Methodology}
The proposed Turvey-Shapley value LRMC methodology involves interleaving of a novel probabilistic approach to the Turvey perturbation method with the SV \textit{characteristic function}, and is illustrated in Algorithm \ref{ShapleyvalueExact1}. 
In this section, we explain the steps for the Turvey LRMC estimation and SV cost allocation.

%\noindent
%\subsubsection{Turvey LRMC estimation} 
%To estimate LRMC using the \textit{Turvey} method, we use the following steps
%\begin{itemize}
First, we calculate the line limit $x$ of the given network with line augmentation cost $Y$. This is given as the yearly peak demand of the network (i.e.~grand coalition of customers) multiplied by a factor of 1.5 (to account for distribution line emergency limit). We have assumed that the probability of a coalition's 50 POE peak demand exceeding the line limit follows a Weibull distribution, $X\sim\mathcal{W}(\alpha,\beta)$\footnote{This choice is not essential to the method; any other fat-tailed distribution could be used as well.}. 
The two-parameter Weibull distribution function is defined as:
\begin{equation}
F_{\alpha,\beta}(x) = 1 - \exp \Big[-\Big(\frac{x}{\alpha}\Big)^{\beta}\Big] \quad \mathrm{for} \quad x \geq 0
\end{equation}  
where $\alpha>0$ is the \textit{scale} parameter and $\beta>0$ is the \textit{shape} parameter. In order to obtain the standard fat-tailed Weibull distribution that is required in this study, $\beta$ is taken as 1.5.

Then, for each coalition, we assume a yearly peak demand growth rate of 1\% as the 50 POE value, which is taken as the mean $\mu$ of the Weibull distribution. Given the mean and shape factor, we calculate the scale factor of the distribution.
%\end{itemize}
% \subsubsection{Incremental cost (IC) calculations}
% \begin{itemize}
If the tail probability $P(X \geq x)$ of a coalition's 50 POE peak demand exceeding the line limit  is less than 0.001, we neglect the coalition cost (set as zero) in the incremental cost (IC) calculations for a particular customer in each coalition size.
For example, if $Y$ is \$1M, then we neglect coalitions with cost less than \$1k, which improves the incremental cost computation time of each customer.
Otherwise, the coalition cost is given as $P(X \geq x) Y$, that is, the coalition's expected LRMC under the corresponding Weibull distribution. 
%\end{itemize}    
%\subsubsection{Shapley value calculation} 
The SV for each customer $i \in \mathcal{N}$ is calculated as the average of its marginal contributions to all coalitions containing $i$, as in \eqref{defShapleyValue1}.
 
% In real terms, if $Y$ is \$~1 million, then we neglect coalitions with marginal cost less than \$~1,000. This step slightly improves the IC computation time of each customer, because we do not need to perform IC calculations at all coalition sizes. 
 
\par In the next three subsections, we describe the exact SV algorithm (\textit{Exact}), and two algorithms (\textit{Sampling} and \textit{Clustering}) to compute the approximate SV for up to 25 customers.

\subsection{Direct Enumeration}
 The exact algorithm (\textit{Exact}), also known as \textit{direct enumeration}, is based on \eqref{defShapleyValue1} and is described in Algorithm \ref{ShapleyvalueExact1}. In terms of computational speed, it performs well with $n<20$ players. But with $n>25$, its performance (w.r.t. speed and memory requirements) deteriorates because of the time taken and memory required to compute the large ($2^{n}-1$) $\times$ $n$ coalition matrix. Note that Line 2 in Algorithm \ref{ShapleyvalueExact1} can be broken down into $n$ coalition sizes according to \eqref{eqn:SV_expanded}. 
% While this alternative approach would lower the memory requirements, it increases the computation time.
 
%  Conversely, in the second exact algorithm (\textit{Exact 2}), described in Algorithm \ref{ShapleyvalueExact2}, \eqref{defShapleyValue2} is broken down to coalition sizes, according to \eqref{eqnSV2}. The advantage of this is that the coalition matrix and respective cost functions are not needed so the incremental costs are computed as needed. In this case, there are lower memory limitations compared to \textit{Exact}.

%\vspace{-1.5em}

%Instead, we do this only for a sufficient number of coalitions, for each coalition size with total coalitions greater than 10,000.

\subsection{Randomised Sampling} \label{section_randomised_sampling}
\par As explained in Section \ref{section_shapley_value}, we use a sample-based randomised algorithm which statistically estimates \eqref{eqn:SV_randomised}, to provide approximate SV calculations based on \eqref{eqn:SV_expanded}. With this, we do not perform all the incremental cost calculations ($2^{n-1}$) required to compute the exact SV for each customer. Instead, we do this only for coalition sets that contain more than 10,000 possible coalitions of the same size. 
In Algorithm \ref{ShapleyvalueApprox} (\textit{Sampling}), we first select randomly a pilot sample ($|\mathcal{P}| = K$) from such large coalitions, where $K$ is determined by trial and error. Then, the standard deviation $\sigma_{\mathcal{P}}$ of the marginal contribution of customer $i$ to the sampled coalition is computed, followed by the calculation of the optimal sample size using:
\begin{equation} \label{eqn6}
    |\mathcal{P}'| = \Big(\frac{Z\sigma_{\mathcal{P}}}{d}\Big)^2
\end{equation}
\noindent where $Z =1.96$ is the z-score of 95\% confidence in a Gaussian distribution and $d =0.2$ is taken to be the margin of error for the sampling estimation.

	\begin{algorithm}[t]
	\caption{Randomised Sampling Algorithm (\textit{Sampling})}\label{ShapleyvalueApprox}
	\small
	%\Procedure{MyProcedure}{}
	
	%\textbf{Inputs:}\\
	$\mathcal{N}$: set of players/customers, $\mathcal{N} = \{1,...,n\}$ \\
	$\mathcal{K}$: partition of set of $k$-sized coalitions, $\mathcal{K} = \{\mathcal{S}^1,..,\mathcal{S}^k,..,\mathcal{S}^{n}\}$ %$(\left\vert\mathcal{S}\right\vert=2^{\left\vert\mathcal{N}\right\vert -1}$)
	%$\mathcal{G}$: set of all coalitions of size $s$ %$g \in \mathcal{G}$ \\
	
\begin{algorithmic}[1]
		\For{each customer $i \in \mathcal{N}$}
		\For{each set of $k$-sized coalitions $\mathcal{S}^k \in \mathcal{K}$}
		\State \textbf{Find} $\mathcal{G}\subset \mathcal{S}^k: \forall\ \mathcal{Z} \in \mathcal{G}, i \in \mathcal{Z}$ %\cap i \ne \{\}$
		\If{$|\mathcal{S}^k| < 10000$}
		\For{each coalition $g \in \mathcal{G}$}
		\State $\theta_g = \frac{(k-1)!\;(n-k)!}{n!}\left(w(g) - w(g - \{i\})\right)$
		\EndFor
		\State $\vartheta_{\mathcal{S}^k} = \sum_{g\in \mathcal{G}}\theta_g$
		\Else
		\State \textbf{Sample} $|\mathcal{P}|$ coalitions from $|\mathcal{G}|:|\mathcal{P}| = K$%0.1|\mathcal{G}|$ 
		\For{each coalition $p \in \mathcal{P}$}
		\State $\theta_p = \left(w(p) - w(p - \{i\})\right)$
		\EndFor
		\State \textbf{Compute} standard deviation $\sigma_{\mathcal{P}}$ of all $\theta_{p \in \mathcal{P}}$
		\State \textbf{Compute} optimal sample size, $|\mathcal{P}'|$ using \eqref{eqn6}
		\State \textbf{Sample} $|\mathcal{P}'|$ coalitions from $|\mathcal{G}|$
		\For{each coalition $p' \in \mathcal{P}'$}
		\State $\theta_{p'} = \frac{(k-1)!\;(n-k)!}{n!}\left(w(p') - w(p' - \{i\})\right)$
		\EndFor
		\State $\vartheta_{\mathcal{S}^k} = \sum_{p'\in \mathcal{P}'}\theta_{p'}$
		\EndIf
		\EndFor
		\State $\phi_i = \sum_{\mathcal{S}^k \in \mathcal{K}}\vartheta_{\mathcal{S}^k}$
		\EndFor
	\end{algorithmic}
\end{algorithm}

 \subsection{Clustering Method}

%(|C|-1)!\;(n-|C|)!
The clustering method illustrated in Algorithm \ref{ShapleyvalueClustering} is also based on \eqref{eqn:SV_expanded}. In this method, we first cluster $|\mathcal{H}^m| = 125$ customers from the Ausgrid \textit{Solar Home Electricity Data} set into $n=5$ representative load profile clusters (with minimum customer set  $|\mathcal{H}^1| = 25$). Here, end-users are clustered based on their half-hourly average daily consumption pattern for a year using the k-means clustering algorithm. Next, for each set of network users $\mathcal{H}^h \in \mathcal{H}$, we find the yearly demand (with 30-minute resolution) of each cluster $i\in \mathcal{N}$, by summing the half-hourly demand of all customers belonging to cluster $i$.
Then, we find the average contribution $\boldsymbol{V}_g^{\mathcal{H}^h_i}$ of member customers to the yearly peak demand of cluster $i$ over all coalitions $g\in\mathcal{G}$. 
%This proportion is later used in allocating the cluster cost to each customer belonging to the cluster.

After computing the SV of each cluster, the cluster cost is then apportioned to its member customers according to their contribution. It is worth noting that the algorithm can be scaled to compute the SV for $|\mathcal{H}^m| > 125$ network users in our dataset, with just an insignificant clustering overhead computation cost for allocating customers into 5 clusters; and moreover, it would scale to settings with up to 25 clusters irrespective of the total number of customers. 

\vspace{-0.6em}

%The clustering procedure is terminated after 10,000 initializations and the best arrangement (best total sum of distances from the centroid) is chosen as the final customer cluster allocation. 

%It is worth noting that the representative load profiles are considered as the centroid of each cluster.
%      % We compute the Shapley value $\mathcal{S}$ of each coalition for each of the 5 clusters.
%       Therefore, the monthly cost of a customer is given as $t\mathcal{S}$
    
%      The yearly Shapley value $SV$ of all customers, which is given as the sum of the monthly Shapley values, are normalised according to:

% $SV^\mathrm{norm} = \frac{SV}{\sum\limits_{i \in \mathcal{N}}SV_{i}}$ 

.   
 % \end{itemize}  
% \begin{figure}[t]
% \centering
% \includegraphics[scale = 0.65]{profiles.pdf}
% %\includegraphics[height=4.8 cm, width=8.5cm]{Figures/figure1}
% \caption{Generic Load Profiles}
% \label{fig4}
% \end{figure}

% \subsection{Shapley value computation}
% \begin{itemize}

%\end{itemize}
    
%\section{Shapley value correlation with peak demand indicators}
% \begin{itemize}

% \item We define the following three peak demand indicators as follows:

% \begin{itemize}
%     \item \textit{Coincident}: This refers to a customer's coincident peak demand based on the network's yearly peak load.
%   \item \textit{Individual}: This refers to an individual customer's yearly peak demand.
%   \item \textit{Total}: This refers to the sum total of a customer's monthly peak demand values in a year.
% \end{itemize}

% \item Next, we find the linear correlation coefficient between the Shapley value and the peak demand indicators.
% \end{itemize}

% $\mathcal{P}(X)$ $\euscr{P}(X)$ $\mathscr{P}(X)$ $\powerset(X)$

%\SetInd{0.25em}{0.05em}
\begin{algorithm}[t]
	\caption{Clustering Algorithm (\textit{Clustering})}\label{ShapleyvalueClustering}
	\small
	%\Procedure{MyProcedure}{}
	
	%\textbf{Inputs:}\\
	$\mathcal{H}$: partition of set of network users, $\mathcal{H}:= \{\mathcal{H}^1,..,\mathcal{H}^h,..,\mathcal{H}^m\}$\\
	$\mathcal{N}$: set of players/clusters, $\mathcal{N} = \{1,...,n\}$ \\
	$\mathcal{K}$: partition of set of $k$-sized coalitions, $\mathcal{K} = \{\mathcal{S}^1,..,\mathcal{S}^k,..,\mathcal{S}^{n}\}$ %$(\left\vert\mathcal{S}\right\vert=2^{\left\vert\mathcal{N}\right\vert -1}$)
	%$\mathcal{G}$: set of all coalitions of size $s$ %$g \in \mathcal{G}$ \\
	
\begin{algorithmic}[1]
    \State \textbf{Cluster} $|\mathcal{H}^m|$ customers into $n$ clusters
    \For{each $\mathcal{H}^h \in \mathcal{H}$}
     \For{$t \longleftarrow 1\ \mathbf{to}\ 100\ \mathbf{step}\ 1$}\Comment{100 MC simulations}
		  \State \textbf{Sample} $|\mathcal{H}^h|$ customers uniformly from $|\mathcal{H}^m|$ 
		\For{each cluster $i \in \mathcal{N}$}
		\For{each set of $k$-sized coalitions $\mathcal{S}^k \in \mathcal{K}$}
		\State \textbf{Find} $\mathcal{G}\subset \mathcal{S}^k: \forall\ \mathcal{Z} \in \mathcal{G}, i \in \mathcal{Z}$ %\cap \{i\} \ne \{\}$
		\For{each coalition $g \in \mathcal{G}$}
		\State \textbf{Find} $\mathcal{H}^h_i\subset \mathcal{H}^h$ \Comment{customers in cluster $i$}
		\State $\boldsymbol{V}_g^{\mathcal{H}^h_i} = [v_g^1,...,v_g^{|\mathcal{H}^h_i|}]$ \Comment{cust. contr. to $w(g)$}
		\State $\theta_g = \frac{(k-1)!\;(n-k)!}{n!}\left(w(g) - w(g - \{i\})\right)$
		\EndFor
		\State $\vartheta_{\mathcal{S}^k} = \sum_{g\in \mathcal{G}}\theta_g$
		\EndFor
		\State $\phi_i = \sum_{\mathcal{S}^k \in \mathcal{K}}\vartheta_{\mathcal{S}^k}$ \Comment{SV of cluster $i$}
		\State $\boldsymbol{\phi_t^{\mathcal{H}^h_i}} = (\phi_i/|\mathcal{G}|)\sum_{g \in \mathcal{G}} \boldsymbol{V}_g^{\mathcal{H}^h_i}$ \Comment{SV--customers in $i$}
		 \EndFor
		\EndFor
		\EndFor
	\end{algorithmic}
\end{algorithm}

% \begin{figure*}[!htb] 
% \centering
% \renewcommand{\thesubfigure}{a}
% 	\subfloat[]{
% 	\includegraphics[scale = 0.64]{Figures/newest_time_taken.pdf} \label{sv_time} 
% 	} \hspace{0.6em}
% 	\renewcommand{\thesubfigure}{b}
% 	\subfloat[]{% 
% 		\includegraphics[scale = 0.64]{Figures/turvey_sv_error.pdf} \label{sv_error}
% 	}
% 	\caption{(a) SV computation time for 5 -- 25 customers (b) Error in Approximate SV Calculation for 5 -- 25 customers}
% 	\label{sv_indices}
% \end{figure*}

% 	\hbox{\hspace{0.6em} \includegraphics[scale = 0.64]{Figures/latest_turvey_sv_error.pdf}}

\begin{figure}[t]
	\centering
	\renewcommand{\thesubfigure}{a}
	\subfloat[]{ \label{sv_time} 
	\hbox{%\hspace{0.2em}
	\includegraphics[scale = 0.68]{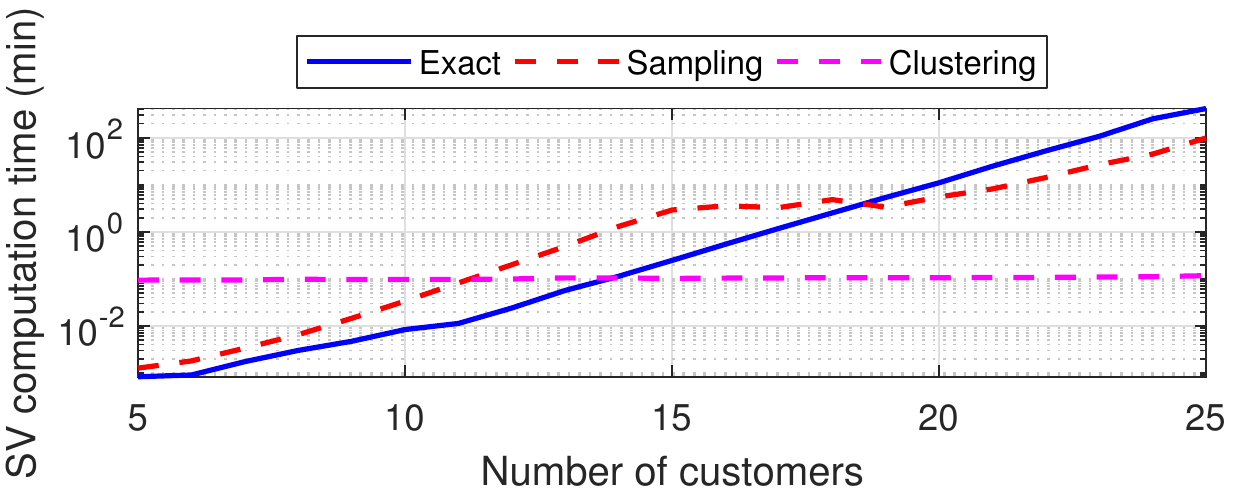}} }
	 \vspace{0.1em}
	\renewcommand{\thesubfigure}{b}
	\subfloat[]{ \label{sv_error}
	\hbox{\hspace{-1.2em} \includegraphics[scale = 0.68]{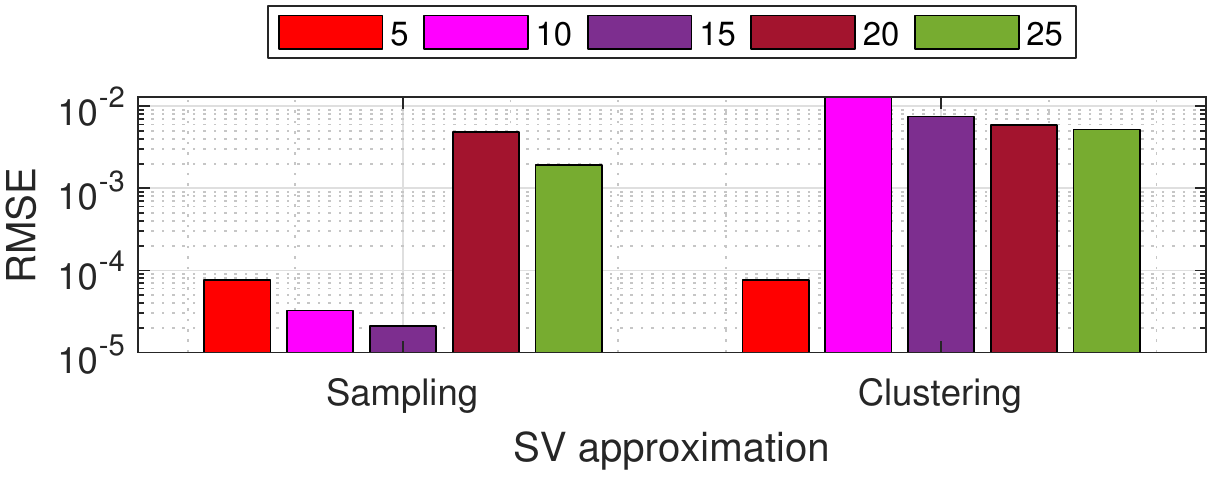}}} 
	\caption{(a) SV computation time for 5 -- 25 customers (b) Error in approximate SV calculation for 5 -- 25 customers. For both plots, the y-axis is in log scale.}
	\label{sv_indices}
\end{figure}

%e provide discussions and analysis of the numerical results obtained from our case studies in the following sub-headings. W
\section{Case Study, Results and discussion} \label{results}
In this section, we assess the computational performance and accuracy of the SV approximation methods, the correlation of SV with three peak demand indicators commonly used to define tariffs, and compare alternative pricing methods with the SV cost allocation. 

To begin, we define the following three peak demand indicators as follows:
\begin{itemize}
    \item \textit{Coincident} peak demand (CPD): This refers to a customer's coincident peak demand at the time of the network's yearly peak load.
   \item \textit{Individual} peak demand (IPD): This refers to a customer's yearly peak demand.
  \item \textit{Total} peak demand (TPD): This refers to the sum total of a customer's monthly peak demand values in a year.
\end{itemize}
% To obtain the above defined values for each customer, we take the absolute value of the net load trace for a year

%For these, we assume two scenarios--customers with and without PV.
%Second

%\subsection{Case Study -- Net Load Traces and Network Tariffs}
\par As case study, the net load traces (solar PV and demand) used in this work were sourced from the Ausgrid (DNSP in NSW) \textit{Solar Home Electricity Data}. The dataset comprises three years of half-hourly resolution smart meter data for the period between July 2010 to June 2013, for 300 residential customers in the Sydney region of Australia. However, we could only extract 125 customers from this dataset with complete solar PV and demand data, for the period between July 2012 to June 2013. This information is used to obtain the above defined peak demand indicators for each customer.

\par We also employ network tariff data from Ausgrid\footnote{Ausgrid Network Price List. Available at https://www.ausgrid.com.au/Industry/Regulation/Network-prices.}, given in Table \ref{table1}\footnote{Peak: 
summer weekdays (Nov. to Mar.) between 2pm to 8pm, winter weekdays (Jun. to Aug.) between 5pm to 9pm; Shoulder: weekdays year round, between 7am to 10pm (exc. Peak periods); Off-peak: all other times.}, which enables us make a rough estimate of the revenue obtained for these customers under the flat and ToU energy network prices.

%Ausgrid Network Price List. Available at 

% \begin{table}[t]
% 	\footnotesize
% 	\centering
% 	\caption{Network Tariff Data}
% 	%\vspace{-1.5em}
% 	\label{table1}
% 	\begin{tabular}[t]{c@{\hspace{0.25cm}}c@{\hspace{0.25cm}}c@{\hspace{0.25cm}}c@{\hspace{0.25cm}}c@{\hspace{0.25cm}}c@{\hspace{0.2cm}}}
% 		\hline 
% 		%& & & \textbf{Network} & & & \\ \hline
% 		\multicolumn{1}{c}{\begin{tabular}[c]{@{}c@{}}Tariff\\ Type \end{tabular}} & \begin{tabular}[c]{@{}c@{}}Fixed\\ charge\\ \si{c/day}\end{tabular} & \begin{tabular}[c]{@{}c@{}}Anytime\\ Energy\\ \si{c/kWh}\end{tabular} & 
% 		\begin{tabular}[c]{@{}c@{}}Off peak\\ Energy\\ \si{c/kWh}\end{tabular} & \begin{tabular}[c]{@{}c@{}}Shoulder\\ Energy\\ \si{c/kWh}\end{tabular} & \begin{tabular}[c]{@{}c@{}}Peak\\ Energy\\ \si{c/kWh}\end{tabular} & %\begin{tabular}[c]{@{}c@{}}Demand\\ Charge\\ \si{\$/kW/month}\end{tabular} \\ 
% 		\hline
% %		Flat & 0.8568 & 11.0321 & - & - & - & - \\ 
% %		ToU & 0.8568 & - & 4.6287 & 12.6922 & 13.9934 & - \\ 
% %		FlatD & 0.8568 & 3.2169 & - & - & - & 4.2112 \\ 
% %		ToUD & 0.8568 & - & 2.1419 & 3.4771 & 4.0804 & 4.2112 \\	\hline
% %		& & & \textbf{Retail} & & & \\ \hline
% 		\textit{Flat} & 40.097 & 11.163 & - & - & -  \\ 
% 		\textit{ToU} & 40.097 & - & 2.805 & 7.086 & 27.335  \\ 
% 		%\textit{FlatD} & 1.5511 & 23.5018 & - & - & - & 4.2112 \\ 
% 		%\textit{ToUD} & 1.5511 & - & 18.8532 & 27.9319 & 28.6750 & 4.2112 
% 		\hline
% 	\end{tabular}
% \end{table}

\begin{table}[t]
	\footnotesize
	\centering
	\caption{Network Tariff Data}
	\vspace{-1.5em}
	\label{table1}
	\begin{tabular}[t]{c@{\hspace{0.25cm}}c@{\hspace{0.25cm}}c@{\hspace{0.25cm}}c@{\hspace{0.25cm}}c@{\hspace{0.25cm}}c@{\hspace{0.2cm}}c@{\hspace{0.1cm}}}
		\hline 
		%& & & \textbf{Network} & & & \\ \hline
		\multicolumn{1}{c}{\begin{tabular}[c]{@{}c@{}}Tariff\\ Type\end{tabular}} & \begin{tabular}[c]{@{}c@{}}Fixed\\ charge\\ \si{c/day}\end{tabular} & \begin{tabular}[c]{@{}c@{}}Anytime\\ Energy\\ \si{c/kWh}\end{tabular} & 
		\begin{tabular}[c]{@{}c@{}}Off peak\\ Energy\\ \si{c/kWh}\end{tabular} & \begin{tabular}[c]{@{}c@{}}Shoulder\\ Energy\\ \si{c/kWh}\end{tabular} & \begin{tabular}[c]{@{}c@{}}Peak\\ Energy\\ \si{c/kWh}\end{tabular} & \\ \hline %\begin{tabular}[c]{@{}c@{}}Feed-in\\ Tariff\\ \si{c/kWh}\end{tabular} 
		\textit{Flat} & 40.097 & 11.163 & - & - & - \\ 
		\textit{ToU} & 40.097 & - & 2.805 & 7.086 & 27.335 \\ \hline
	\end{tabular}
\end{table}

\vspace{-0.5em}

\subsection{SV Computation Time and Accuracy} 
\par Here, we compare the computational performance and accuracy of the different SV calculation algorithms. For this first set of computations, we have assumed that all customers possess solar PV, so the net load is used to compute their monthly peak demand. Fig. \ref{sv_time} shows the SV computation time in minutes for all customer sizes from 5 up to 25 users. A related point to consider is that the exact SV computation for more than 25 users is not computationally feasible. The \textit{Exact} algorithm performs best for $n < 15$ customers, but with $n \geq 15$, this is not the case. It takes 418 minutes to compute the SV for $n=25$ customers, due to the time and memory consuming coalition matrix generation and the corresponding coalition cost function calculations.

% However, \textit{Exact 2} takes 17 times the time to perform same computation for $n = 25$ customers. The advantage of splitting the coalition into $n$ coalition sizes in \textit{Exact 2} is only to overcome the memory limitations of \textit{Exact}.
%, but it performs better than \textit{Exact} for $n \geq 25$.

% \par Contrarily, \textit{Sampling 1} takes 1,120 minutes to compute the SV for $n=25$ (more than twice the time for \textit{Exact}). This is due to the computational requirements of the set membership test in the \textit{Sampling 1} algorithm (Line 3). More so, since the the cost function calculation is done multiple times in \textit{Sampling 1} compared to a single calculation in \textit{Exact}, the approximate algorithm is adversely affected. This is because the time taken for the search algorithm to find a coalition's yearly peak demand grows linearly with the size of the search space. Nonetheless, \textit{Sampling 1} is expected to outperform \textit{Exact} for very large number of players.

\par Conversely, there is a significant improvement in computational performance with the first approximate algorithm compared to \textit{Exact}. \textit{Sampling} takes only 98 minutes to compute the SV for $n=25$ (about a quarter of the time taken for \textit{Exact}). This reduction in computation time is as a result of performing IC calculations for a select (optimal) sample, using a constant number as a pilot sampling size, instead of performing the IC calculations for all coalitions (in \textit{Exact}) at the same time. Furthermore, splitting the total coalitions into $n$ coalition sizes in \textit{Sampling} overcomes the memory limitations of \textit{Exact}. %For \textit{Approx. 3}, however, the reduction in computation time is by chance and is not guaranteed for $n>25$. 
%taking a percentage of the coalitions in each coalition size as the pilot sampling size
 \par On the other hand, the clustering algorithm takes the least time to compute the SV for $n \geq 15$ customers. This is because the SV calculation is done for only 5 players (or clusters), with a little overhead computation cost for clustering. 

\par To evaluate the accuracy of the sampling and the clustering technique, we find the root mean square error (RMSE) in the SV estimation. Fig. \ref{sv_error} shows RMSE values between $10^{-2}$ and $10^{-5}$ relative to mean values between $0.04$ and $0.2$ for $n\leq25$ customers. Although the sampling approach is more accurate than the clustering technique for $5<n\leq25$, it cannot be up-scaled to $n>25$ players, without a significant increase in the computation time. Besides, with the clustering technique, the estimation error reduces as more customers are added to make the clusters more representative.

%exact SV and the approximate SVs obtained using \textit{Approx. 1 to 3}. On evaluating the accuracy of the approximate algorithms by taking  Figure \ref{fig2}
%%%%%%%%%%%%%%%%%%%%%%%%%%%%%%%%%%%%%%%%%%%%%%%%%%%%%%%%%%%%%%%%%%%%%%%%%%%
% \begin{figure*}[!htb] 
% \centering
% \renewcommand{\thesubfigure}{a}
% 	\subfloat[]{
% 	\includegraphics[scale = 0.64]{Figures/correlation_nopv_yearly.pdf} \label{sv_correlation_nopv} 
% 	} \hspace{1.2em}
% 	\renewcommand{\thesubfigure}{b}
% 	\subfloat[]{% 
% 		\includegraphics[scale = 0.64]{Figures/correlation_pv_yearly.pdf} \label{sv_correlation_pv}
% 	}
% 	\caption{(a) SV linear correlation coefficient without PV (b) SV linear correlation coefficient with PV}
% 	\label{sv_correlation}
% \end{figure*}

\begin{figure}[t]
	\centering
	\renewcommand{\thesubfigure}{a}
	\subfloat[Without PV]{ \label{sv_correlation_nopv}
	\includegraphics[scale = 0.68]{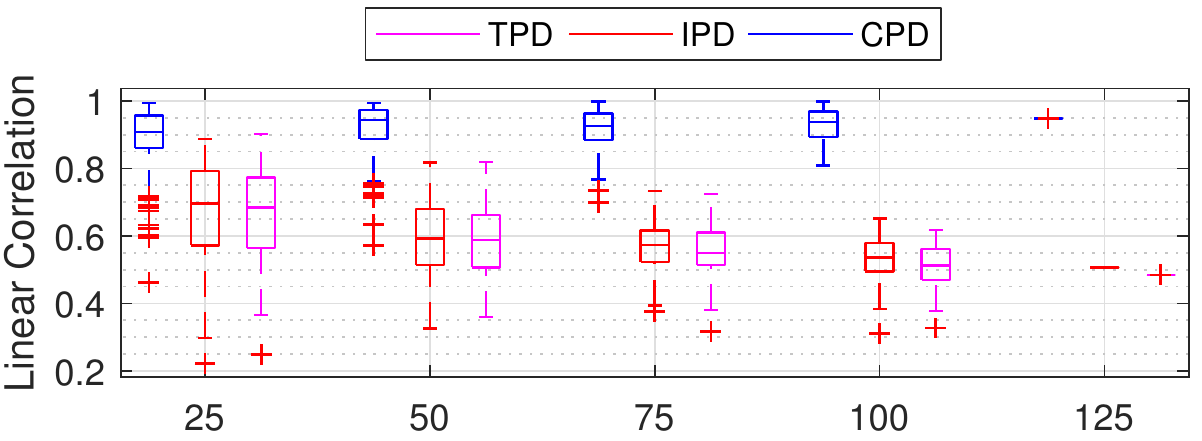}} 
	 \vspace{0.1em}
	\renewcommand{\thesubfigure}{b}
	\subfloat[With PV]{\label{sv_correlation_pv}
	\hbox{\hspace{-0.5em} \includegraphics[scale = 0.68]{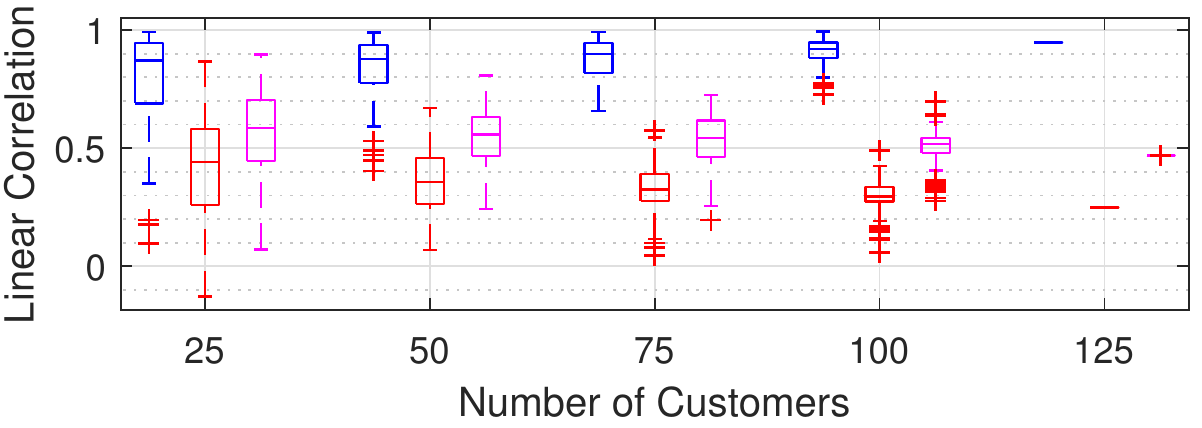}}} 
	\caption{SV linear correlation coefficient box plot.  CPD--coincident peak demand, IPD--individual peak demand, TPD--total peak demand. Note that there is no statistical variance for $n=125$.}
	\label{sv_correlation}
\end{figure}

%%%%%%%%%%%%%%%%%%%%%%%%%%%%%%%%%%%%%%%%%%%%%%%%%%%%%%%%%%%%%%%%%%%%%%%%%%%

% \begin{figure*}[!htb] 
% \centering
% %\captionsetup{justification=centering}
% 	\hspace{-0.8em}
% 	\subfloat{% 
% 		\includegraphics[scale = 0.6]{Figures/Coincident_mean_corr_nopv_yearly.pdf}  %\label{figure5a}
% 	}  \hspace{0.1em}
% 	\renewcommand{\thesubfigure}{a}
% 	\subfloat[SV linear correlation scatter plot Without PV]{% 
% 		\includegraphics[scale = 0.6]{Figures/Individual_mean_corr_nopv_yearly.pdf} %\label{figure5b}
% 	} \hspace{0.2em}
% 	\subfloat{% 
% 		\includegraphics[scale = 0.6]{Figures/Total_mean_corr_nopv_yearly.pdf} %\label{figure5b}
% 	} 
	
% 	\subfloat{% 
% 		\includegraphics[scale = 0.6]{Figures/Coincident_mean_corr_pv_yearly.pdf} \label{figure5c}
% 	} \hspace{0.1em}
% 	\renewcommand{\thesubfigure}{b}
% 	\subfloat[SV linear correlation scatter plot with PV]{% 
% 		\includegraphics[scale = 0.6]{Figures/Individual_mean_corr_pv_yearly.pdf} \label{sv_correlation_peak_nopv}
% 	} \hspace{0.2em}
% 	\subfloat{% 
% 		\includegraphics[scale = 0.6]{Figures/Total_mean_corr_pv_yearly.pdf} \label{sv_correlation_peak_pv}
% 	}
% 	\caption{SV linear correlation scatter plot for 125 customers}
% 	\label{sv_correlation_peak}
% \end{figure*}

\begin{figure}[t] 
\centering
	%\hspace{0.001em}
	\subfloat{% 
		\includegraphics[scale = 0.64]{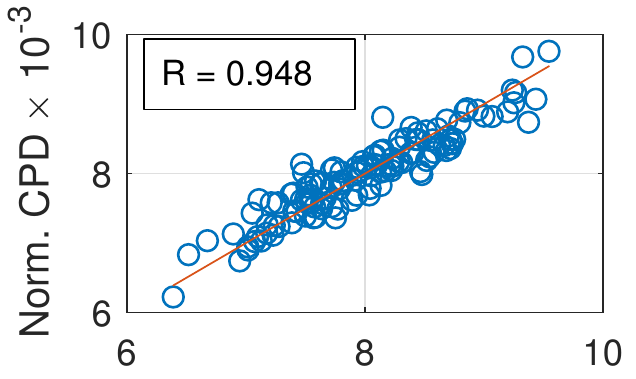}  
	}  \hspace{0.1em}
	\subfloat{% 
		\includegraphics[scale = 0.64]{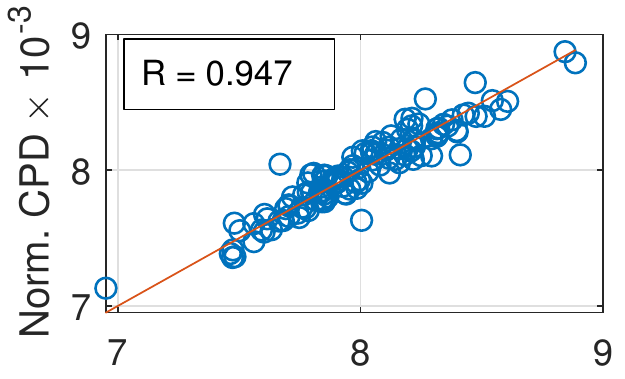} 
	} 
	\vspace{-0.15em}
	\subfloat{% 
	\hbox{\hspace{-0.2em}	\includegraphics[scale = 0.64]{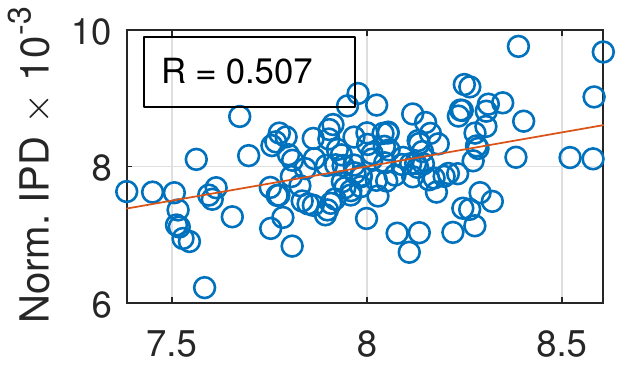}} 
	} \hspace{0.2em}
	\subfloat{% 
		\includegraphics[scale = 0.64]{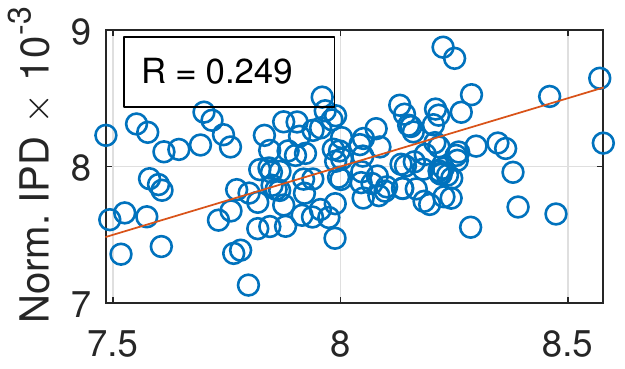} 
	} 
	\renewcommand{\thesubfigure}{a}
	\subfloat[Without PV]{ \label{sv_correlation_peak_nopv}
	\hbox{\hspace{-0.6em} \includegraphics[scale = 0.64]{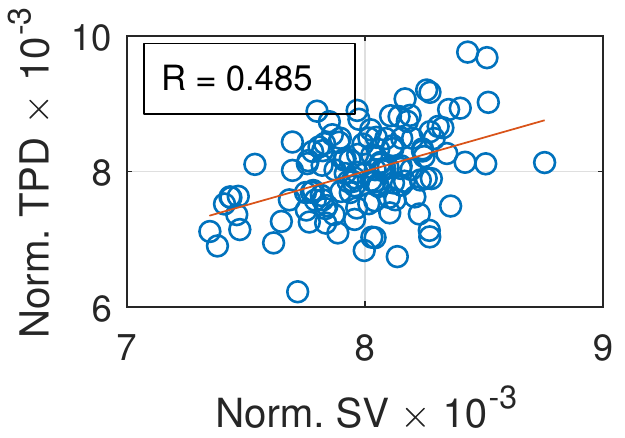}}  
	} \hspace{-0.5em}
	\renewcommand{\thesubfigure}{b}
	\subfloat[With PV]{ \label{sv_correlation_peak_pv} 
		\includegraphics[scale = 0.64]{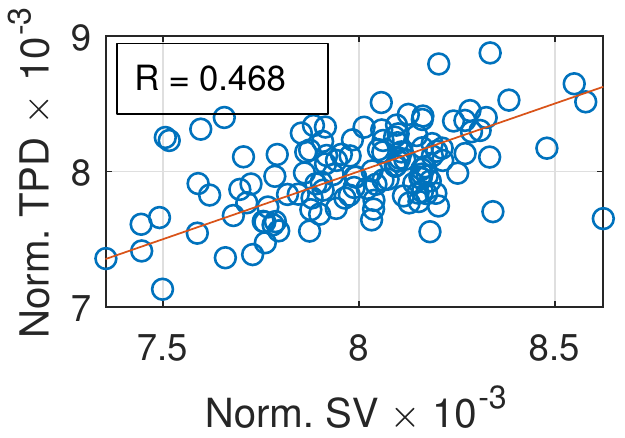} 
	}
	\caption{SV linear correlation scatter plot for 125 customers. CPD--coincident peak demand, IPD--individual peak demand, TPD--total peak demand.}
	\label{sv_correlation_peak}
\end{figure}

\begin{table}[t]
\footnotesize
\centering
\caption{Mean SV correlation with Peak demand Indicators}
\begin{tabular}{c@{\hspace{0.14cm}}c@{\hspace{0.14cm}}c@{\hspace{0.3cm}}c@{\hspace{0.3cm}}c@{\hspace{0.3cm}}c@{\hspace{0.3cm}}c@{\hspace{0.2cm}}}
\hline
\multirow{2}{*}{Scenario} & \multirow{2}{*}{\begin{tabular}[c]{@{}c@{}}Peak demand\\ Indicator\end{tabular}} & \multicolumn{5}{c}{Customers} \\ 
 &  & 25 & 50 & 75 & 100 & 125 \\  \hline
\multirow{3}{*}{Without PV} & Coincident & 0.8773 & 0.9134 & 0.9114 & 0.9272 & 0.9478 \\ 
 & Individual & 0.6668 & 0.5891 & 0.5644 & 0.5321 & 0.5065 \\  
 & Total & 0.6626 & 0.5833 & 0.5541 & 0.5140 & 0.4848 \\ \hline
\multirow{3}{*}{With PV} & Coincident & 0.7967 & 0.8380 & 0.8780 & 0.9098 & 0.9473 \\ 
 & Individual & 0.4113 & 0.3625 & 0.3291 & 0.2918 & 0.2493 \\  
 & Total & 0.5700 & 0.5512 & 0.5277 & 0.5016 & 0.4685 \\ \hline
\label{sv_corr}
\end{tabular}
\end{table}

\vspace{-1em}

% 	\caption{(a) Correlation between cost allocation method and peak demand variants without PV (top) and with PV (bottom) (b) Error in cost allocation from the optimal without PV (top) and with PV (bottom)}

%In Figure \ref{fig3}, we show the linear relationship between the exact Shapley value  and the pre-defined peak demand indicators .

%simulation runs
\subsection{SV Linear Correlation with Peak demand Indicators}
This section shows the results obtained by finding the linear correlation between the SV computed using the \textit{clustering} technique and the peak demand indicators, for two scenarios (i) all customers without PV and (ii) all customers with PV. Since the SV is computed for 100 Monte Carlo runs based on uniform random sampling, the Pearson's correlation coefficients (R-value) are presented as box plots in Fig. \ref{sv_correlation} while Table \ref{sv_corr} shows the mean values, for $\mathcal{H}$ of size 25, 50, 75, 100, and 125.

% described in Section \ref{method}
\par For Scenario 1 (Fig. \ref{sv_correlation_nopv}), the SV correlates more with \textit{Individual} peak demand than with \textit{Total} peak demand but for Scenario 2 (Fig. \ref{sv_correlation_pv}), the converse is the case. It is worth noting that \textit{Individual} and \textit{Total} corresponds to charging customers based on their yearly peak load and monthly peak load respectively. Without PV, a customer's true demand is revealed, which is less sensitive to weather, so individual peak demand dominates. However, with PV, a customer's demand profile is modified with PV generation which is season-dependent, and as such it's better to charge customers on a monthly basis. Nevertheless, for both scenarios, the SV correlates most with \textit{Coincident} peak demand, because it drives augmentation cost the most. 

Figs. \ref{sv_correlation_peak_nopv} and \ref{sv_correlation_peak_pv} show the scatter plot for SV correlation with the peak demand indicators for $n=125$ customers without PV and with PV respectively. For Scenario 1, the mean R-values are 0.948, 0.507 and 0.485, for \textit{Coincident}, \textit{Individual}, and \textit{Total} peak demand respectively while the R-values for Scenario 2 are 0.947, 0.249 and 0.468, for \textit{Coincident}, \textit{Individual}, and \textit{Total} peak demand respectively. While CPD and TPD have similar values in both scenarios, IPD is considerably different. This is because a customer's (individual) yearly peak demand changes significantly with the addition of PV.

%For both scenarios, the R-values for \textit{Coincident} and \textit{Total} peak demand are relatively close, but considerably different for \textit{Individual} peak demand since a customer's yearly peak demand changes significantly with the addition of PV.

%This implies that for very small number of players/customers, there is a possibility that the Shapley value will correlate more with \textit{Total} and \textit{Individual} peak demand, than with the \textit{Coincident} peak demand as aggregate or individual demand dominates. In reality, a distribution network consists of customers well beyond $n=5$ and coincident peak demand becomes dominant.

% \begin{table}[t]
% \footnotesize
% \centering
% \caption{Median SV correlation with Peak demand Indicators}
% \begin{tabular}{c@{\hspace{0.12cm}}c@{\hspace{0.12cm}}ccccc}
% \hline
% \multirow{2}{*}{} & \multirow{2}{*}{\begin{tabular}[c]{@{}c@{}}Peak demand\\ Indicator\end{tabular}} & \multicolumn{5}{c}{Customers} \\ 
%  &  & 25 & 50 & 75 & 100 & 125 \\  \hline
% \multirow{3}{*}{No PV} & Coincident & 0.9079 & 0.9443 & 0.9254 & 0.9375 & 0.9478 \\
%  & Individual & 0.6955 & 0.5919 & 0.5730 & 0.5366 & 0.5065 \\  
%  & Total & 0.6841 & 0.5879 & 0.5498 & 0.5126 & 0.4848 \\ \hline
% \multirow{3}{*}{With PV} & Coincident & 0.8707 & 0.8772 & 0.8985 & 0.9207 & 0.9473 \\ 
%  & Individual & 0.4413 & 0.3577 & 0.3251 & 0.2955 & 0.2493 \\ 
%  & Total & 0.5841 & 0.5584 & 0.5436 & 0.5166 & 0.4685 \\ \hline
% \end{tabular}
% \end{table}

%%%%%%%%%%%%%%%%%%%%%%%%%%%%%%%%%%%%%%%%%%%%%%%%%%%%%%%%%%%%%%%%%%%%%%%%%%%
%\begin{figure*}[!htb] 
\begin{figure}[t] 
\centering
	%\hspace{-0.2em}
	\subfloat{% 
		\includegraphics[scale = 0.66]{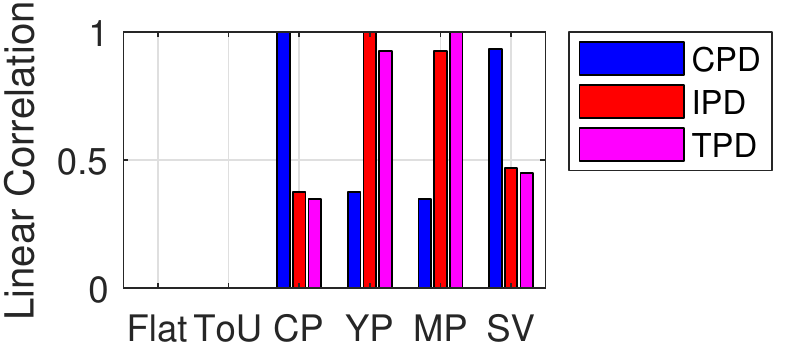}  \label{figure5a}
	}  \hspace{-0.6em}
	\subfloat{% 
		\includegraphics[scale = 0.66]{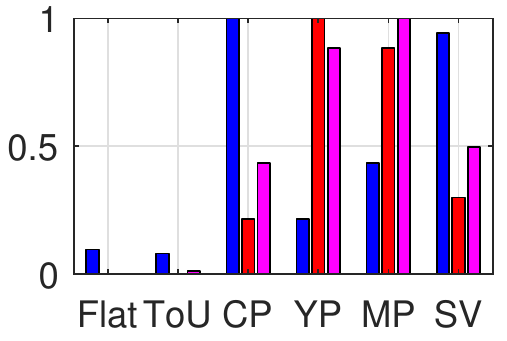} \label{figure5b}
	} 
	\vspace{0.5em}
	\renewcommand{\thesubfigure}{a}
	\subfloat[Without PV]{ \label{sv_alloc} 
	\hbox{\hspace{-0.5em}\includegraphics[scale = 0.65]{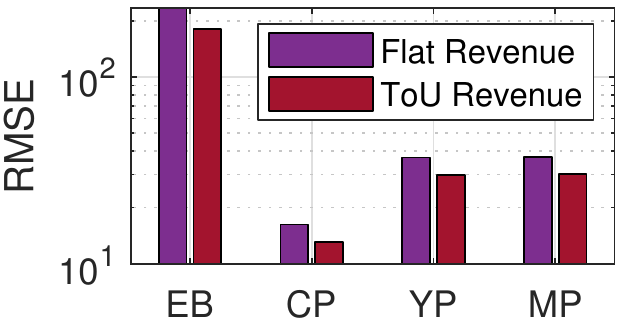}} 
	} %\hspace{0.8em}
	\renewcommand{\thesubfigure}{b}
	\subfloat[With PV]{\label{sv_alloc_error}
		\hbox{\hspace{0.7em}	\includegraphics[scale = 0.65]{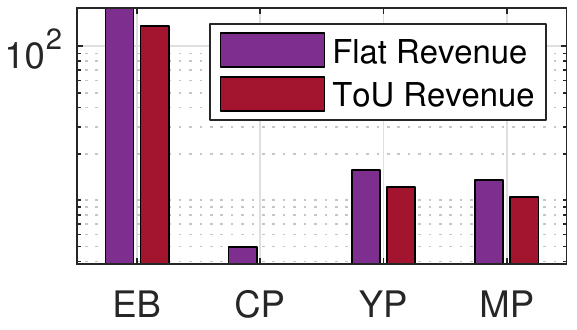} }
	}
	\caption{(a) Correlation between cost allocation method and peak demand variants (top) and error in cost allocation from the SV allocation (bottom). EB--energy-based, CP--coincident peak, YP--yearly peak, MP--monthly peak.}
	\label{sv_corr_error}
\end{figure}
%%%%%%%%%%%%%%%%%%%%%%%%%%%%%%%%%%%%%%%%%%%%%%%%%%%%%%%%%%%%%%%%%%%%%%%%%%%
\vspace{-0.2em}

\subsection{Comparison of Cost Allocation Methodologies}
Here, we evaluate how different energy-based and peak demand-based cost allocation methodologies compare to the SV, by measuring the correlation between normalised customer cost allocation and specifically-defined peak demand indicators. First, we perform the SV computation for $n=125$ customers, for both scenarios. Then, we estimate the revenue obtained by the DNSP for these customers under the two tariffs (Flat- and ToU-based) described in Section \ref{method}. For this, we have have neglected the feed-in-tariff (FiT) as it is administered through a different mechanism and handled by retailers in the Australian electricity system's regulatory and billing arrangements. Therefore, we use only the power import from the grid for our calculations. 
\par We consider the following cost allocation methods in our analysis:

\begin{itemize}
%\item Optimal cost allocation (Shapley value)
\item Energy-based (EB - Flat or ToU)
\item Coincident peak load (CP - \textit{Coincident} peak)
\item Yearly peak load (YP - \textit{Individual} peak)
\item Monthly peak load (MP - \textit{Total} peak) and
\item Shapley value cost allocation (SV)
\end{itemize}

In the first method, cost allocation is done according to the revenue calculations under Flat and ToU tariff for $n=125$ customers, using tariff values in Table \ref{table1}. For the rest, we split the total revenue obtained using the energy-based tariffs, according to the normalised SV and the normalised \textit{Coincident}, \textit{Individual}, and \textit{Total} peak demand values for each customer.

%Flat-based and ToU-based revenue respectively %the two scenarios
The top part of Fig. \ref{sv_corr_error} shows the correlation between the different cost allocation methodologies and the peak demand indicators. From these, we can deduce that the SV provides a fine balance between coincident, individual and aggregate peak demand, since it properly accounts for network usage at times other than the coincident network peak. This is by virtue of the way it is being computed, by evaluating the marginal contribution of customers to all possible coalitions. Although, the major cost driver for distribution networks is the coincident peak demand, it is necessary for a cost-reflective tariff design to appropriately account for aggregate and individual customer peak demand. Furthermore, the yearly and monthly peak demand allocation are also not cost-reflective, since they have a lower correlation with coincident peak demand compared with the SV allocation. Energy-based allocation methods perform worst as they show much lower correlation with the peak demand indicators. However, the results for Scenario 2 show that ToU-energy based allocation is better than that of Flat-energy based allocation as it shows comparatively higher R-values. Moreover, these results implicitly show that inter-customer subsidies will be reduced since customers would be paying their fair share, given the similar R-value for \textit{Coincident} peak demand in both scenarios.

\par At the bottom part of Fig. \ref{sv_corr_error}, we show the error (RMSE) in actual cost values that will arise when a DNSP allocates cost to customers using less optimal cost allocation methodologies. This translates to inaccurate wealth transfer amongst customers in a distribution network. As expected, there are higher RMSE values resulting from energy-based cost allocation methods. This implies that they are least cost-reflective for both scenarios, both in terms of cost-causality and equity in cost allocation. Also, for both scenarios, the coincident peak load allocation results in the least error because the main cost driver for networks is the coincident peak. While the monthly peak allocation, and the yearly peak demand allocation method results in similar RMSE values for Scenario 1, this is not the case for the second scenario. When all customers possess PV, the monthly peak allocation results in slightly lower RMSE compared to the yearly peak demand allocation. This shows that for residential customers with PV, using a monthly peak demand network tariff is more cost reflective than a yearly peak demand tariff. It will be unfair to charge customers based on their sole highest yearly peak demand, which occurs in just one month of a year, and does not account for the seasonality in energy consumption which comes with PV generation.

%\vspace{-0.5em}

\section{Conclusions and Future Work} \label{conclusions}
\par In this work, we showed the efficacy of the \textit{Turvey-Shapley value} method in calculating and apportioning LRMC to network users in a cost-reflective way and with low computational burden using the proposed clustering technique. 

\par We have demonstrated that the SV is a cost-reflective cost allocation method for networks with a large number of customers, regardless of PV adoption. This is because the SV is a principled technique, which provides a proper balance between the peak demand indicators (network cost drivers). This makes for a fair cost allocation, which further reduces inter-customer subsidies. Other peak demand-based cost allocation approaches perform well up to the extent to which they appropriately balance the peak demand indicators, but with a greater emphasis on coincident peak demand. Furthermore, our results show that energy-based cost allocation methodologies are least cost-reflective as they least correlate with the peak demand indicators.

\par For future work, we will consider LRMC allocation for customers with both PV and batteries. In this case, an optimisation would have to be solved for each cost function computation.

% \par The downside of the Shapley value is its computational performance for large number of players, which is typical for most real world applications. Therefore, for future work, we will use better sampling approaches to compute the Shapley value for a significantly higher number of customers in a distribution network. More so, we aim to carry out a probabilistic analysis of the Shapley value in the computation of the coalition cost using the \textit{Turvey method}. Here, we consider a pre-determined network cost (or long-run marginal cost) which relates to the network capacity limit. This means that the cost of coalition follows a distribution and customers in any coalition will be charged depending on the probability of their aggregate consumption approaching the network line capacity. 

\bibliographystyle{IEEEtran}
\small{\bibliography{turvey_shapley}}

\begin{thebibliography}{10}
\providecommand{\url}[1]{#1}
\csname url@samestyle\endcsname
\providecommand{\newblock}{\relax}
\providecommand{\bibinfo}[2]{#2}
\providecommand{\BIBentrySTDinterwordspacing}{\spaceskip=0pt\relax}
\providecommand{\BIBentryALTinterwordstretchfactor}{4}
\providecommand{\BIBentryALTinterwordspacing}{\spaceskip=\fontdimen2\font plus
\BIBentryALTinterwordstretchfactor\fontdimen3\font minus
  \fontdimen4\font\relax}
\providecommand{\BIBforeignlanguage}[2]{{%
\expandafter\ifx\csname l@#1\endcsname\relax
\typeout{** WARNING: IEEEtran.bst: No hyphenation pattern has been}%
\typeout{** loaded for the language `#1'. Using the pattern for}%
\typeout{** the default language instead.}%
\else
\language=\csname l@#1\endcsname
\fi
#2}}
\providecommand{\BIBdecl}{\relax}
\BIBdecl

\bibitem{picciariello2015distributed}
A.~Picciariello, J.~Reneses, P.~Frias, and L.~S{\"o}der, ``Distributed
  generation and distribution pricing: Why do we need new tariff design
  methodologies?'' \emph{Electric Power Systems Research}, vol. 119, pp.
  370--376, 2015.

\bibitem{rubio2000marginal}
F.~J. Rubio-Od{\'e}riz and I.~J. Perez-Arriaga, ``Marginal pricing of
  transmission services: A comparative analysis of network cost allocation
  methods,'' \emph{IEEE Trans. Power Systems}, vol.~15, no.~1, pp. 448--454,
  2000.

\bibitem{zolezzi2001review}
J.~Zolezzi, H.~Rudnick, F.~Danitz, J.~Bialek, J.~Pan, Y.~Teklu, S.~Rahman, and
  K.~Jun, ``Review of usage-based transmission cost allocation methods under
  open access [discussion],'' \emph{IEEE Trans. Power Systems}, vol.~16, no.~4,
  pp. 933--934, 2001.

\bibitem{sotkiewicz2006allocation}
P.~M. Sotkiewicz and J.~M. Vignolo, ``Allocation of fixed costs in distribution
  networks with distributed generation,'' \emph{IEEE Trans. Power Systems},
  vol.~21, no.~2, pp. 639--652, 2006.

\bibitem{li2008cost}
F.~Li, N.~P. Padhy, J.~Wang, and B.~Kuri, ``Cost-benefit reflective
  distribution charging methodology,'' \emph{IEEE Trans. Power Systems},
  vol.~23, no.~1, pp. 58--64, 2008.

\bibitem{brown2015efficient}
T.~Brown, A.~Faruqui, and L.~Grausz, ``Efficient tariff structures for
  distribution network services,'' \emph{Economic Analysis and Policy},
  vol.~48, pp. 139--149, 2015.

\bibitem{bonbright1961principles}
J.~C. Bonbright, A.~L. Danielsen, and D.~R. Kamerschen, \emph{Principles of
  public utility rates}.\hskip 1em plus 0.5em minus 0.4em\relax Columbia
  University Press New York, 1961.

\bibitem{abdelmotteleb2018designing}
I.~Abdelmotteleb, T.~G{\'o}mez, J.~P.~C. {\'A}vila, and J.~Reneses, ``Designing
  efficient distribution network charges in the context of active customers,''
  \emph{Applied Energy}, vol. 210, pp. 815--826, 2018.

\bibitem{lewis1941two}
W.~A. Lewis, ``The two-part tariff,'' \emph{Economica}, vol.~8, no.~31, pp.
  249--270, 1941.

\bibitem{boiteux1952determination}
M.~P. Boiteux and P.~Stasi, ``{Sur la d{\'e}termination des prix de revient de
  d{\'e}veloppement dans un syst{\'e}me interconnect{\'e} de
  production-distribution},'' {International Union of Producers and
  Distributors of Electrical Energy (UNIPEDE)}, Tech. Rep., 1952.

\bibitem{nijhuis2017analysis}
M.~Nijhuis, M.~Gibescu, and J.~Cobben, ``Analysis of reflectivity \&
  predictability of electricity network tariff structures for household
  consumers,'' \emph{Energy Policy}, vol. 109, pp. 631--641, 2017.

\bibitem{passey2017designing}
R.~Passey, N.~Haghdadi, A.~Bruce, and I.~MacGill, ``Designing more cost
  reflective electricity network tariffs with demand charges,'' \emph{Energy
  Policy}, vol. 109, pp. 642--649, 2017.

\bibitem{nelson2013new}
T.~Nelson and F.~Orton, ``A new approach to congestion pricing in electricity
  markets: Improving user pays pricing incentives,'' \emph{Energy Economics},
  vol.~40, pp. 1--7, 2013.

\bibitem{picciariello2015electricity}
A.~Picciariello, C.~Vergara, J.~Reneses, P.~Fr{\'\i}as, and L.~S{\"o}der,
  ``Electricity distribution tariffs and distributed generation: Quantifying
  cross-subsidies from consumers to prosumers,'' \emph{Utilities Policy},
  vol.~37, pp. 23--33, 2015.

\bibitem{simshauser2014network}
P.~Simshauser, ``Network tariffs: resolving rate instability and hidden
  subsidies,'' \emph{Working paper 45, AGL Applied Economic and Policy
  Research}, 2014.

\bibitem{pimm2018time}
A.~J. Pimm, T.~T. Cockerill, and P.~G. Taylor, ``Time-of-use and time-of-export
  tariffs for home batteries: Effects on low voltage distribution networks,''
  \emph{Journal of Energy Storage}, vol.~18, pp. 447--458, 2018.

\bibitem{ahmad2018pricing}
A.~Faruqui, ``Pricing directions: A stakeholder perspective,'' {Submission in
  response to the NSW DNSPs 2019-24 regulatory proposals and AER issues paper},
  Tech. Rep., 2018.

\bibitem{ahmad2018rate}
------, ``Rate design 3.0 -- future of rate design,'' \emph{Public Utilities
  Fortnightly}, May 2018.

\bibitem{stenner2015australian}
K.~Stenner, E.~Frederiks, E.~V. Hobman, and S.~Meikle, ``{Australian consumers'
  likely response to cost-reflective electricity pricing},'' CSIRO Australia,
  2015.

\bibitem{turvey1969marginal}
R.~Turvey, ``Marginal cost,'' \emph{The Economic Journal}, vol.~79, no. 314,
  pp. 282--299, 1969.

\bibitem{Shapley1953}
L.~S. Shapley, ``A value for n-person games,'' in \emph{Contributions to the
  Theory of Games}, H.~W. Kuhn and A.~W. Tucker, Eds.\hskip 1em plus 0.5em
  minus 0.4em\relax Princeton University Press, 1953, vol.~28, pp. 307--317.

\bibitem{ChalkiadakisEtal2011}
G.~Chalkiadakis, E.~Elkind, and M.~Wooldridge, \emph{Computational Aspects of
  Cooperative Game Theory (Synthesis Lectures on Artificial Inetlligence and
  Machine Learning)}, 1st~ed.\hskip 1em plus 0.5em minus 0.4em\relax Morgan \&
  Claypool Publishers, 2011.

\bibitem{biggardarr}
D.~Biggar, ``{An exploration of NERA's proposed approach to estimating long-run
  marginal cost},'' {Sapere Research Group Limited}, Tech. Rep., 27 January
  2012.

\bibitem{neraconsult}
{NERA Economic Consulting}, ``{Estimating Long Run Marginal Cost in the
  National Electricity Market -- A Paper for the AEMC},'' AEMC, Tech. Rep., 19
  December 2011.

\bibitem{toothrichard}
R.~Tooth, ``Measuring long run marginal cost for pricing,'' Sapere Research
  Group Limited, Tech. Rep., March 2014.

\bibitem{o2015shapley}
G.~O'Brien, A.~El~Gamal, and R.~Rajagopal, ``{Shapley value estimation for
  compensation of participants in demand response programs},'' \emph{IEEE
  Trans. Smart Grid}, vol.~6, no.~6, pp. 2837--2844, 2015.

\bibitem{bakr2015using}
S.~Bakr and S.~Cranefield, ``{Using the Shapley Value for Fair Consumer
  Compensation in Energy Demand Response Programs: Comparing Algorithms},'' in
  \emph{Data Science and Data Intensive Systems (DSDIS), 2015 IEEE
  International Conference on}.\hskip 1em plus 0.5em minus 0.4em\relax IEEE,
  2015, pp. 440--447.

\bibitem{Chapman_IREP2017}
A.~C. Chapman, S.~Mhanna, and G.~Verbi\v{c}, ``{Cooperative game theory for
  non-linear pricing of load-side distribution network support},'' in
  \emph{2017 IREP Symposium Bulk Power System Dynamics and Control}, Aug 2017.

\bibitem{zolezzi2002transmission}
J.~M. Zolezzi and H.~Rudnick, ``Transmission cost allocation by cooperative
  games and coalition formation,'' \emph{IEEE Trans. Power Systems}, vol.~17,
  no.~4, pp. 1008--1015, 2002.

\bibitem{tan2002application}
X.~Tan and T.~Lie, ``{Application of the Shapley value on transmission cost
  allocation in the competitive power market environment},'' \emph{IEE
  Proceedings-Generation, Transmission and Distribution}, vol. 149, no.~1, pp.
  15--20, 2002.

\bibitem{khare2015shapley}
S.~Khare, B.~Khan, and G.~Agnihotri, ``{A Shapley value approach for
  transmission usage cost allocation under contingent restructured market},''
  in \emph{2015 International Conference on Futuristic Trends on Computational
  Analysis and Knowledge Management (ABLAZE)}.\hskip 1em plus 0.5em minus
  0.4em\relax IEEE, 2015, pp. 170--173.

\bibitem{sharma2017loss}
S.~Sharma and A.~Abhyankar, ``{Loss allocation for weakly meshed distribution
  system using analytical formulation of Shapley value},'' \emph{IEEE Trans.
  Power Systems}, vol.~32, no.~2, pp. 1369--1377, 2017.

\bibitem{ghassemi2008cooperative}
F.~Ghassemi and V.~Krishnamurthy, ``A cooperative game-theoretic measurement
  allocation algorithm for localization in unattended ground sensor networks,''
  in \emph{2008 11th International Conference on Information Fusion}.\hskip 1em
  plus 0.5em minus 0.4em\relax IEEE, 2008.

\bibitem{stanojevic2010economic}
R.~Stanojevic, N.~Laoutaris, and P.~Rodriguez, ``{On economic heavy hitters:
  Shapley value analysis of 95th-percentile pricing},'' in \emph{Proceedings of
  the 10th ACM SIGCOMM conference on Internet measurement}.\hskip 1em plus
  0.5em minus 0.4em\relax ACM, 2010, pp. 75--80.

\bibitem{byun2009fair}
S.-S. Byun, H.~Moussavinik, and I.~Balasingham, ``{Fair allocation of sensor
  measurements using shapley value},'' in \emph{IEEE 34th Conference on Local
  Computer Networks}.\hskip 1em plus 0.5em minus 0.4em\relax IEEE, 2009, pp.
  459--466.

\bibitem{fatima2008linear}
S.~S. Fatima, M.~Wooldridge, and N.~R. Jennings, ``{A linear approximation
  method for the Shapley value},'' \emph{Artificial Intelligence}, vol. 172,
  no.~14, pp. 1673--1699, 2008.

\bibitem{fatima2007randomized}
------, ``{A randomized method for the Shapley value for the voting game},'' in
  \emph{Proceedings of the 6th international joint conference on Autonomous
  agents and multiagent systems}.\hskip 1em plus 0.5em minus 0.4em\relax ACM,
  2007, p. 157.

\bibitem{david2005shapley}
L.~David, O.~Massol, and A.~Moison, ``{The Shapley value as joint cost
  allocation mechanism: is the story definitely over},'' \emph{URL citeseerx.
  ist. psu. edu/viewdoc/summary}, 2005.

\bibitem{castro2009polynomial}
J.~Castro, D.~G{\'o}mez, and J.~Tejada, ``{Polynomial calculation of the
  Shapley value based on sampling},'' \emph{Computers \& Operations Research},
  vol.~36, no.~5, pp. 1726--1730, 2009.

\bibitem{maleki2013bounding}
S.~Maleki, L.~Tran-Thanh, G.~Hines, T.~Rahwan, and A.~Rogers, ``{Bounding the
  estimation error of sampling-based Shapley value approximation},''
  \emph{arXiv preprint arXiv:1306.4265}, 2013.

\bibitem{azuatalam2019shapley}
D.~Azuatalam, G.~Verbi\v{c}, and A.~Chapman, ``{Shapley value analysis of
  distribution network cost-causality pricing},'' in \emph{{2019 PowerTech
  Conference, Milan}}.\hskip 1em plus 0.5em minus 0.4em\relax IEEE, 2019.

\end{thebibliography}
\end{document}